\documentclass[useAMS,usenatbib]{mn2e}
\usepackage{graphicx}
\usepackage{times}

\title[Lithium in the symbiotic Mira V407\,Cyg]{Lithium in the symbiotic
Mira V407\,Cyg}
\author[Tatarnikova et al.]{A.A. Tatarnikova$^{1}$, P.M. Marrese$^{2}$, U. 
Munari$^{2}$, T. Tomov$^{3}$, P. A. Whitelock$^{4}$, B.F. Yudin$^{1}$\\
$^{1}$Sternberg Astronomical Institute,Moscow, Russia\\
$^{2}$INAF-Osservatorio Astronomico di Padova, Sede di Asiago, 36012 Asiago 
(VI), Italy\\
$^{3}$Centre for Astronomy, Nicholaus Copernicus University, ul. Gagarina 11, 
87-100 Torun, Poland\\
$^{4}$South African Astronomical Observatory, PO Box 9, Observatory 7935, South
Africa, email:paw@saao.ac.za}
\begin{document}

\date{Accepted 12 June 2003. Received \ldots; in original form \ldots}

\pagerange{\pageref{firstpage}--\pageref{lastpage}} \pubyear{2003}

\maketitle

\label{firstpage}

\begin{abstract}
We report an identification of the lithium resonance doublet LiI
6708\,\AA\ in the spectrum of V407\,Cyg, a symbiotic Mira with a
pulsation period of about 745 days. The resolution of the spectra used was $R
= \lambda/\Delta\lambda \approx 18500$ and the measured equivalent width of
the line is $\sim 0.34$\,\AA.  It is suggested that the lithium enrichment
is due to hot bottom burning in the intermediate mass AGB variable, although
other possible origins cannot be totally ruled out. In contrast to
lithium-rich AGB stars in the Magellanic clouds, ZrO 5551\,\AA, 6474\,\AA\
absorption bands were not found in the spectrum of V407\,Cyg. These are the
bands used to classify the S-type stars at low-resolution. Although we
identified weak ZrO 5718\,\AA, 6412\,\AA\ these are not visible in the
low-resolution spectra, and we therefore classify the Mira in V407\,Cyg as
an M type. This, together with other published work, suggests lithium
enrichment can precede the third dredge up of s-process enriched material in
galactic AGB stars.
\end{abstract}

\begin{keywords}
Stars: AGB and post-AGB -- Stars: variable -- (stars:) binaries: symbiotic
-- stars: individual: V407 Cyg -- stars: chemically peculiar 
\end{keywords}

\section{Introduction}
\subsection{Lithium in AGB stars}

Lithium is one of the key elements related to stellar evolution and
thermonuclear synthesis. It is both easily destroyed and easily produced at
the temperatures found in stellar interiors. Lithium undergoes nuclear
burning at temperatures above about $2.5 \times 10^6$\,K and primordial
lithium is therefore destroyed in the convective atmospheres of late-type
giants. Fresh lithium can be produced and brought to the surface of AGB
stars via the beryllium transport mechanism proposed by \citet{CF}. While
the detailed determination of lithium abundances in AGB star atmospheres is
difficult, due to heavy line blanketing and non-LTE effects, AGB stars with
unusually high abundances of lithium (lithium rich $\log \varepsilon
({\rm Li}) >1.0$ and super lithium rich (SLR) $\log \varepsilon
({\rm Li}) >4.0$ stars) are known both in the Galaxy and in the
Magellanic Clouds; although the two environments seem to produce lithium
enrichment in different types of AGB star (see below).

In the Magellanic Clouds \cite{Set} found 35 lithium-rich stars. The
majority of these are S-type stars while a small fraction are C stars; the
paucity of C-type SLR stars in the LMC compared to the Galaxy was recently
confirmed by \cite{Hat}. Most of the lithium-rich stars in the Magellanic
Clouds lie in the luminosity range $-6>M_{bol}>-7$ mag and their enrichment
can be explained as a consequence of hot bottom burning (HBB), in stars with
initial masses of $4<M_{\odot}<6$, as described by \cite{SB}. These SLR
stars have rather thin dust shells (possibly a selection effect) and those
which are also large amplitude variables have periods over 400 days and are
more luminous than the Mira period-luminosity relation would predict
(Whitelock et al. 2003).

A rather small number of SLR S (e.g. RZ Sgr, T Sgr, VX Aql, He 166) and C
(e.g. T Ara, IY Hya, WZ Cass, WX Cyg) stars are known in the Galaxy, a much
larger fraction of them being C stars than in the Magellanic Clouds and some
are J-type C stars (i.e. they have high $^{13}$C/$^{12}$C); see Catchpole \&
Feast (1971,1976) and \cite{Aet}.  The distances, and therefore the
luminosities, of these galactic SLR stars are uncertain, but the evidence
favours their evolving from lower mass progenitors than those of the HBB
stars in the Magellanic Clouds. \cite{AI} discuss possible mechanisms for
mixing lithium to the surface in low mass stars, but a full explanation for
their enrichment remains elusive.

\cite{G-Let} have reported preliminary results from a
lithium survey of IRAS selected AGB variables which shows that lithium is
preferentially found in stars with long periods, $P>400$\,d, large expansion
velocities, $V>6$\,km\,s$^{-1}$, and relatively thin dust shells,
$[12]-[25]<0$ (although this is largely a selection effect as most of those
with thicker shells could not be observed spectroscopically). Furthermore
these are all oxygen-rich, although not S-type stars, and could be related
to the HBB stars in the Magellanic Clouds.  However, insufficient
information is presented to judge their luminosities and or the level of
their lithium enrichment.

Lithium is also found in the close binary systems HD 172481 and HD 190390
(Reyniers 2002) where its origin is not understood. 

In this note we report on the LiI 6708\,\AA\ doublet identification in
the spectrum of the symbiotic Mira V407\,Cyg.

\subsection{Symbiotic Miras}

 The symbiotic Miras are binary systems in which an asymptotic giant branch
variable interacts with a white dwarf; these are rare objects, fewer than 28
are known in the Galaxy and LMC (Whitelock 1987, 2003). The AGB star looses
mass to the white dwarf via its stellar wind, while a circumbinary nebulae
reaches ionization levels comparable to those found in planetary nebulae.
Their binary periods are assumed to be long, in excess of 10 years, while
the pulsation periods of the AGB stars range from 275 to 745 days. 

V407\,Cyg was identified as symbiotic Mira by Munari, Margoni \& Stagni
(1990); at 745 days its pulsation period is the longest known in a symbiotic
Mira. Most symbiotic Miras suffer from strong and variable circumstellar
extinction. V407\,Cyg is an exception, it has surprisingly low circumstellar
extinction given its long period and symbiotic character. According to the
intensity of the TiO 7055\,\AA\ band the Mira can be classified as M6-M7. VO
absorption bands are also present in its spectrum. Therefore the cool
component of V407\,Cyg is oxygen rich (Kolotilov et al. 1998).

Note that if we accept that the late-type giant in V407\,Cyg is a Mira
variable then we must also accept that it is an AGB star. Mira variables
obey a period-luminosity relation (e.g. Whitelock et al. 2003 and references
therein) and even the short period Miras, found in globular clusters, have
luminosities above the tip of the first giant branch (e.g. Feast \&
Whitelock 1987).

 The kinematics of AGB variables shows that their pulsation periods are a
function of the population to which they belong, with longer period stars
being younger (e.g. Feast \& Whitelock 2000). Not much is known about stars
with periods over 700 days, but by analogy with Olivier, Whitelock \& Marang
(2001) we can estimate that an AGB star on the period-luminosity relation
with a period of 745 days would have had a progenitor of around
2\,$M_{\odot}$. If, however, it is undergoing HBB then its initial mass
could be quite different and presumably larger than this. Normal Miras, i.e.
those not undergoing HBB, obey a period-luminosity relation which has proved
useful for determining distances within the Galaxy. HBB Miras are
over-luminous (Whitelock et al. 2003) and the PL relation will therefore
underestimate their distances.

\section{Lithium identification}

High-resolution spectra ($R = \lambda/\Delta\lambda \approx 18500$) of
V407\,Cyg have been obtained with the Echelle spectrograph mounted at the
Cassegrain focus of the 1.82m telescope which is operated by the Padova and
Asiago Astronomical Observatories on Mt. Ekar (Asiago, Italy). The detector
was a Thomson THX31156 CCD with 1024$\times$1024 19 $\mu$m pixels, and the
slit width was set to 1.8 arcsec. The spectra cover the 4500-9000\,\AA\
range. The reduction and analysis of the spectra was performed in the
standard fashion under IRAF.

\begin{figure}
\includegraphics[angle=-90,width=84mm]{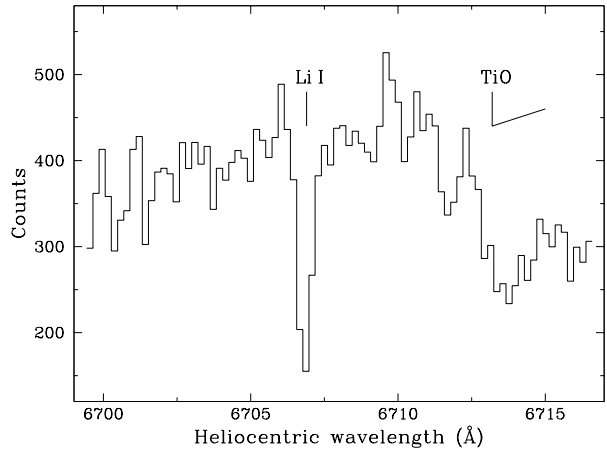}
\caption{The region around the LiI 6708\,\AA\ line in the spectrum of 
V407\,Cyg obtained on 30 July 1997.}
\label{fig:fig1}
\end{figure}

Figure\,1 illustrates a fraction of the V407\,Cyg spectrum around the line 
we suspect to be lithium.
Mindful of the possible confusion of the LiI doublet at
6707.76\,\AA\ and 6707.91\,\AA\ with CeII at 6708.099\AA\, as discussed by
Reyniers et al. (2002), the wavelength of the feature was
checked in detail. The radial velocity of V407\,Cyg was established for
three spectra by cross-correlating a 200\,\AA\ continuum strip in the
vicinity of the line in question with a similar region from HD\,18884, an
M2III radial velocity standard with $V=-25.8$\,km\,s$^{-1}$; a few telluric
lines were masked out. The measurements are accurate to about
$0.2$\,km\,s$^{-1}$. Table~1 details the velocity derived for V407\,Cyg from
the continuum (column 3), and from the feature assuming it originates from
the blended LiI doublet at 6707.812\,\AA\ (column 4) (Smith, Lambert \&
Nissen 1993), or from CeII (column 5). Column 6 shows the phase of the Mira,
where phase 0.0 corresponds to maximum brightness in the $R_C$ band
according to the ephemeris from Munari \& Jurdana-\v{S}epi\'c (2002).  The
radial velocities will, of course, vary according to the pulsation phase.

\begin{table}
\begin{center}
\caption[]{\label{vels} Radial velocities for V407\,Cyg}
\begin{tabular}{crcccc}
\hline
      sp.no. &   \multicolumn{1}{c}{date}  &    
RV(star) &  RV(LiI)&  RV(CeII) & Mira \\               
& &\multicolumn{3}{c}{{($\rm km\,s^{-1}$)}}& phase\\
\hline                                                                                
       25577 &  30 July 97  &   --40.7    &    --43.7   &   --56.5& 0.120 \\                 
       32948 &   8 June 99  &   --47.1    &    --49.9   &   --62.7& 0.030\\               
       37219 &   9 Sept 01  &   --38.0    &    --40.6   &   --53.3& 0.136\\                 
\hline
\end{tabular}
\end{center}
\end{table}

The table shows a constant difference of $\rm 3.0\,km\,s^{-1}$ between the
radial velocity of the Mira and of the line, if the line is produced by the
blended $\rm ^7Li$ doublet (assuming a negligible contribution from 
$\rm ^6Li$). If, however, the line is identified with CeII the
difference in radial velocity with the Mira is $\rm 15.6\,km\,s^{-1}$. This
indicates that the LiI identification is to be preferred over CeII. The
absence of any strong s-process lines in the spectrum (see below) also
supports this conclusion.

The equivalent width of LiI 6708\,\AA\ is about 0.34\,\AA, 
relative to the line-blanketed local continuum. This is in 
the spectrum obtained on 30 July 1997, when the hot component was in an
inactive state and the equivalent width of the H$\alpha$ emission line was
only $\sim 3$\,\AA. The intensities of the absorption bands TiO 5448\,\AA\
and 7125\,\AA\ suggest a spectral type of M6-M7. Therefore, it is clear
that, at least in this spectral region, the V407\,Cyg radiation was
dominated by the cool component.  On other dates the hot component was in an
active state which is reflected in a reduced equivalent width for the
lithium line.  Although, an intrinsic variability of the lithium line in the
Mira atmosphere, depending, for instance, on the pulsation phase, cannot be
excluded.

A rough estimate of the abundance of lithium may be obtained from the ratio
of the equivalent widths of LiI 6708\,\AA\ to CaI 6573\,\AA\ (see Catchpole
(1982) for a detailed discussion). The equivalent width of the calcium line
in the spectrum of V407\,Cyg is $\sim 0.17$\AA. The equivalent width of the
lithium line in V407\,Cyg, 0.34\,\AA, is approximately the same as
\cite{Boe} observed for the S stars: R\,CMi, SU\,Mon, R\,Cyg and HR\,8714.
Although the ratio of the equivalent width of LiI 6708\,\AA\ to CaI
6573\,\AA\ in V407\,Cyg is about two which is slightly higher than in these
late giants - possibly indicating a higher lithium abundance in V407\,Cyg. 
A similar equivalent width for the lithium line is also found, for instance,
in the S star HV\,5584 in the LMC. Its effective temperature is $\sim
3200$\,K, i.e. approximately that expected for an M6 giant. \citet{Set}
estimated the lithium abundance for HV\,5584 as $\log \varepsilon (Li)
\approx 2.8$.

% i.e. in excess of cosmological abundance derived from the surface abundance
%in population II field stars, $\log \varepsilon (^{7}{\rm Li}) \approx
%2.2\pm0.1$ (Bonifacio \& Molaro 1997).

\begin{figure}
\includegraphics[angle=-90,width=84mm]{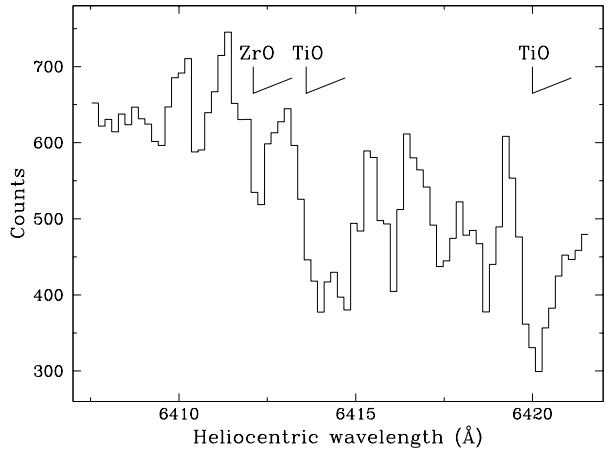}
\caption{The region around the ZrO absorption band in the spectrum of V407\,Cyg 
obtained on 30 July 1997.}
\label{fig:fig2}
\end{figure}

V407\,Cyg is not an S star: the absorption bands of the ZrO molecule at
6474\,\AA\ and 5751\,\AA\ are not visible in our spectra. These bands arise
from the ground state of the ZrO molecule and are used to classify S type
stars in low-resolution spectra \citep{Kee}.  In this context it should be
noted that the nearby TiO band 6479\,\AA\ is relatively strong and is
clearly visible in the low-resolution spectra of V407\,Cyg.  It is worth
emphasizing that the TiO absorption band is observed while ZrO is not,
because at low-resolution it can be difficult to be certain which molecule
is present.

In high-resolution spectra of the symbiotic star CH\,Cyg, which contains an
M7 giant, \citet{J-S} found weak ZrO bands at 5718\,\AA, 6412\,\AA, but not
at 5551\,\AA, 6474\,\AA. In the spectrum of V407\,Cyg at 5718\,\AA\ and
6412\,\AA\ weak absorption features are also present (Figure\,2). Following
\citet{J-S} we can identify these as ZrO bands. In the spectrum of CH\,Cyg
absorptions due to ionized rare-earth elements BaII and LaII are seen, in
particular, the lines BaII 4334\,\AA\ and LaII 4827\,\AA. These absorption
lines are also visible in the spectrum of V407\,Cyg. 

Thus V407\,Cyg does show some weak enhancements of s-process elements which
probably indicates that dredge-up has started.  There also seems to be a
close similarity between the spectra of the cool components in the symbiotic
binaries V407\,Cyg and CH\,Cyg, which have similar spectral types, although
the giant in CH\, Cyg is not a Mira. There is, however, no lithium line in
the spectrum of CH\,Cyg.

%{\bf How do you get the bolometric mag? We don't know the distance and can't
%assume the Mira PL if it is HBB!  Our estimation of its maximum bolometric
%stellar magnitude is $M_{\rm bol} \approx - 4.8$ (the giant's brightness is
%remarkable variable). This value is out of the interval in which according
%to the model calculations the HBB phase begins.}

\section{Discussion}
 A strong lithium LiI 6708\,\AA\ line has been identified for the first time
in a symbiotic Mira. It is only the second identification of lithium in a
Galactic Mira with a period of over 700 days, \cite{G-Let} having already
found lithium in an unnamed variable of comparable period.

 There are three possible origins for the lithium which are discussed
briefly below. First it could have been produced by the hot star
when it was on the AGB, and deposited onto its companion during the
mass-loss process which left it as a white dwarf. This would require the hot
component to have undergone its AGB to white dwarf transition within about
$10^4$ years of the present epoch; this being the time scale on which the
lithium would be destroyed as convection takes surface material close to the
high temperature core. This seems much less probable than the alternative
explanations explored below.

Secondly, the lithium might have been produced more recently, as a product of
the binary interaction, and deposited onto the surface of the AGB star.
Arnould \& N{\o}rgaard (1975) and Starrfield et al. (1978) predict the
formation of large quantities of lithium during hydrogen-burning
thermonuclear runaways, and Hernanz et al. (1996) have predicted significant
lithium production from some nova models, although it is not at all clear
that these calculations will be applicable to symbiotic systems such as
V407\,Cyg. As mentioned in section 1.1 lithium is detected in binary systems
where its presence is not understood.

Thirdly, we suggest that V407\,Cyg is actually an intermediate mass star,
and that the lithium originates from hot bottom burning, as it does in the
long period S-type Miras in the Magellanic Clouds. Although we cannot rule
out other mechanisms until our understanding of the production of lithium in
galactic low mass AGB stars and in interacting binaries is on a much firmer
footing. We also note that, if our explanation is correct, the surface
enrichment of lithium apparently precedes the dredge of s-process elements
in V407\,Cyg as it does for the AGB variables discussed by \cite{G-Let}.

 This discovery that a significant fraction of long period IRAS Miras are
lithium rich \citep{G-Let}, together with finding lithium in an example of
the rare class of symbiotic Miras, suggests that surveys for lithium should
be extended to more long period variables, both in the Galaxy and in
the Magellanic Clouds.

\section*{Acknowledgments}
This research was partly supported by RFBR grant No. 02-02-16235 and KBN 
research grant No. 5 P03D 003 20, and by ASI grant I-R-050/02. We are
grateful to an anonymous referee for his/her suggestions which improved the
paper.

\bsp
\label{lastpage}
\end{document}